\documentclass[pra,aps,twocolumn,showpacs]{revtex4}
\usepackage{amsfonts,amssymb,eucal}
\usepackage{dcolumn}
\usepackage[decmulti]{inputenc}
\usepackage[centertags]{amsmath}
\usepackage{graphicx}
\usepackage{times}
\usepackage{color}
\begin{document}
\bibliographystyle{aip}
\date{\today}

\newcommand{\coul}{\ensuremath{\text{coul}}}
\newcommand{\erf}{\ensuremath{\text{erf}}}
\newcommand{\erfgau}{\ensuremath{\text{erfgau}}}
\newcommand{\erfc}{\ensuremath{\text{erfc}}}

\newcommand{\lr}{\ensuremath{\text{lr}}}
\newcommand{\sr}{\ensuremath{\text{sr}}}
\newcommand{\csr}{\ensuremath{\text{sr}}}
\newcommand{\HF}{\ensuremath{\text{HF}}}
\newcommand{\KS}{\ensuremath{\text{KS}}}
\renewcommand{\H}{\ensuremath{\text{H}}}
\newcommand{\Rys}{\ensuremath{\text{Rys}}}
\newcommand{\LDA}{\ensuremath{\text{LDA}}}
\newcommand{\unif}{\ensuremath{\text{unif}}}
\newcommand{\res}{\ensuremath{\text{res}}}
\newcommand{\CI}{\ensuremath{\text{CI}}}
\newcommand{\RSH}{\ensuremath{\text{RSH}}}
\newcommand{\MP}{\ensuremath{\text{MP}}}

\newcommand{\Tr}{\ensuremath{\text{Tr}}}

\newcommand{\bra}[1]{\ensuremath{\langle #1 \vert}}
\newcommand{\ket}[1]{\ensuremath{\vert #1  \rangle}}
\newcommand{\braket}[2]{\ensuremath{\langle  #1 \vert #2  \rangle}}
\newcommand{\ketbra}[1]{\ensuremath{\vert{#1}\rangle\langle{#1}\vert}}

\renewcommand{\l}{\ensuremath{\lambda}}
\newcommand{\s}{\ensuremath{\sigma}}
\newcommand{\R}{\ensuremath{\mathbb{R}}}
\renewcommand{\O}[1]{\ensuremath{\mathnormal{O}(#1)}}
\renewcommand{\b}[1]{\ensuremath{\mathbf{#1}}}
\newcommand{\sumN}[1]{\ensuremath{\sum_{#1=1}^N}}
\newcommand{\Psit}{\ensuremath{\tilde{\Psi}}}
\newcommand{\Psitl}{\ensuremath{\tilde{\Psi}^{\lambda}}}
\newcommand{\Psil}{\ensuremath{\Psi^{\lambda}}}
\newcommand{\Psimul}{\ensuremath{\Psi^{\mu,\lambda}}}
\newcommand{\nmul}{\ensuremath{n^{\mu,\lambda}}}
\newcommand{\mul}{\ensuremath{\mu,\lambda}}
\newcommand{\Psik}[1]{\ensuremath{\Psi^{(#1)}}}
\newcommand{\Psimuk}[1]{\ensuremath{\Psi^{\mu,(#1)}}}
\newcommand{\Psimu}{\ensuremath{{\Psi^{\mu}}}}
\newcommand{\Phimu}{\ensuremath{\Phi^{\mu}}}
\newcommand{\m}{\ensuremath{\text{min}}}
\markright{}

\title{van der Waals forces in  density functional theory: \\
perturbational long-range electron interaction corrections}
\author{J\'anos~G.~\'Angy\'an, Iann C.~Gerber}
\affiliation{Laboratoire de Cristallographie et de Mod\'elisation des Mat\'eriaux
Min\'eraux et Biologiques, UMR 7036, CNRS - Universit\'e Henri Poincar\'e,
\\ B.P. 239, F-54506 Vand{\oe}uvre-l\`es-Nancy, FRANCE}

\author{Andreas Savin, Julien Toulouse}
\affiliation{Laboratoire de Chimie Théorique, UMR 7616,
CNRS - Universit\'e Pierre et Marie Curie,\\ 4 place Jussieu, F-75005 Paris, FRANCE}


\begin{abstract}
~\\
Long-range exchange and correlation effects, responsible for the failure of
currently used approximate density functionals in describing van der Waals forces,
are taken into account explicitly after a separation of the electron-electron
interaction in the Hamiltonian into  short- and  long-range components.
We propose a "range-separated hybrid" functional
based on a local density approximation for the short-range
exchange-correlation energy, combined with a long-range exact exchange energy.
Long-range correlation effects are added by a second-order perturbational
treatment. The resulting scheme is general and is particularly
well-adapted to describe van der Waals complexes, like rare gas dimers.
\end{abstract}


\maketitle

\section{Introduction}
\label{sec:intro}

Van der Waals (dispersion) interactions are universal attractive forces
due to long-range correlation of electrons between weakly- or
non-overlapping electron groups~\cite{Dobson:02}. They play an important role in the
cohesive energy of practically all kinds of materials: intermolecular
complexes, extended systems, like molecular crystals, liquids or
biological macromolecules. Although, in principle, density functional theory
(DFT)~\cite{HohKoh-PR-64} within the Kohn-Sham (KS) scheme ~\cite{KohSha-PR-65}
is able to provide the exact ground state energy of an electronic system,
present approximate density functionals are
inappropriate to describe long-range electron correlation and consequently
fail for van der Waals interactions, manifested by their incapacity of reproducing the
correct $R^{-6}$ asymptotic behaviour of the intermolecular potential~\cite{Kristyan:94}.

Several propositions have been published recently to add the missing
long-range correlation contribution or to use asymptotically correct
correlation energy expressions in DFT. Most of these methods require a
partitioning of the system into interacting parts and are valid only for
large separations~\cite{Engel:00,Jansen:03,Misquitta:03}.
Seamless dispersion energy functionals~\cite{Dobson:99},
that are valid for the whole range of possible intermolecular
separations have also been proposed~\cite{Dion:04}. A general problem in
schemes that use an additive correction to standard functionals is
the double counting of a part of the correlation effects that are
already present in the original functional.

Moreover, it is not enough to add missing correlation effects to
traditional density functionals. Many of the present approximate functionals,
like the local density approximation (LDA) which is well-known for
its notorious overbinding tendency, and also many popular generalized
gradient approximations (GGA), already predict a more-or-less
pronounced bound state for simple van der Waals complexes, like rare gas
dimers~\cite{Patton:97}. As it has been pointed out by
Harris twenty years ago~\cite{Harris:85}, this behaviour
is related to the erroneous distance dependence of approximate
exchange functionals. In effect, in self-interaction corrected calculations
the minimum on the potential curve disappears~\cite{Engel:00}.
Therefore, in order to describe correctly both the minimum
and the asymptotic region of van der Waals potential energy surfaces
it is mandatory to remove the unphysical bonding by appropriately correcting
the exchange functional.

Here, we propose a scheme based on a long-range/short-range decomposition
of the electron interaction which meets the above requirements and
remedies the description of van der Waals forces in the framework of a
first-principles approach, which takes into account simultaneously long-range
correlation and exchange effects, avoids double counting and is size-extensive.

Our scheme is based on the hypothesis that for the description of
van der Waals (London) dispersion forces one should improve the
representation of long-range electron interaction (exchange and correlation) effects.
At a first level of approximation, we treat the
long-range exchange energy explicitly while maintaining a density
functional approximation for the short-range
exchange-correlation energy. This step defines a
"range-separated hybrid" (RSH) scheme, which is corrected
in a second step for the long-range correlation effects
by a second-order perturbation theory, leading to
size extensive M{\o}ller-Plesset (MP2)-like correction.
This method will be referred to as RSH+MP2.

The idea of a long-range/short-range decomposition of the electron interaction
is not new (see,
e.g.,~Refs.~\onlinecite{NozPin-PR-58,KohHan-JJJ-XX,Stoll:85,Panas:95,Adamson:96}).
In the context of DFT, this approach has been used to construct
multi-determinantal extensions of the KS
scheme~\cite{SavFla-IJQC-95,Sav-INC-96a,Sav-INC-96,LeiStoWerSav-CPL-97,%
PolSavLeiSto-JCP-02,SavColPol-IJQC-03,PedJen-JJJ-XX,Toulouse:04a}. A density
functional scheme with correct asymptotic behaviour has been proposed along
these lines very recently by Baer and Neuhauser~\cite{Baer:05} and the correct
$1/r$ asymptotic behaviour of the long-range exact exchange has been also
exploited in time-dependent DFT calculations of
polarizabilities~\cite{Ikura:01,Tawada:04}, constituting the major motivation
of the recent development of "Coulomb attenuated" hybrid
functionals~\cite{Yanai:04}. Heyd, Scuseria and Ernzerhof applied an inverse
range-separation in order to get rid of the convergence problems of the exact
exchange in solid-state calculations~\cite{Heyd:03,Heyd:04a}. Their HSE03
functional is a generalization of the PBE0 hybrid functional~\cite{Adamo:99}
where the long-range portion of the exact exchange is replaced by the
long-range component of the PBE exchange functional~\cite{PerBurErn-PRL-96}.

In the context of the calculation of van der Waals energies, the idea of
separating the electron interaction operator
to short- and long-range components has already been explored
by the work of Kohn, Meier and Makarov, who applied the adiabatic
connection -- fluctuation-dissipation approach for long-range
electron interactions~\cite{Kohn:98},  leading to an
asymptotically correct expression of the dispersion forces.
It has also been shown~\cite{Kamiya:02} that
the artificial minimum of the rare gas dimer potential curves
can be removed by an exact treatment of the long-range exchange.

The second order perturbational treatment of the
\textit{full} Coulomb interaction has already been used by
several authors for the van der Waals
problem~\cite{Lein:99,Engel:00,Lotrich:05}, and it was shown that the
resulting asymptotic potential has the qualitatively correct $1/R^6$
form. As shown very recently, quantitatively reliable
asymptotic form of the potential energy curve can be expected
from adiabatic connection -- fluctuation-dissipation theory
calculations~\cite{Furche:05}.

The general theoretical framework is outlined in Section~\ref{sec:theory},
describing the RSH scheme and the
second-order perturbational treatment of long-range correlation effects.
As described in Section~\ref{sec:results}, our approach has been tested
on rare gas dimers. These systems are typical van der Waals complexes, where the
attractive interactions are exclusively due to London dispersion
forces. They constitute a stringent test of the method, since
the potential curves have very shallow minima of the order of
about 100 $\mu$H.

Unless otherwise stated, atomic units is assumed throughout this work.


\section{Theory}
\label{sec:theory}

\subsection{Multi-determinantal extension of the Kohn-Sham scheme}
\label{sec:exact}
We first recall the principle of the multi-determinantal extension
of the KS scheme based on a long-range/short-range decomposition
(see, e.g., Ref.~\onlinecite{Toulouse:04a} and references therein).

The starting point is the decomposition the Coulomb electron-electron
interaction $w_{ee}(r)=1/r$ as
\begin{equation}
 w_{ee}(r) = w_{ee}^{\lr,\mu}(r) +  w_{ee}^{\sr,\mu}(r),
\label{}
\end{equation}
where $w_{ee}^{\lr,\mu}(r)=\erf(\mu r)/r$ is a long-range interaction
and $w_{ee}^{\sr,\mu}(r)$ is the complement short-range interaction.
This decomposition is controlled by a single parameter $\mu$. For $\mu=0$,
the long-range interaction vanishes, $w_{ee}^{\lr,\mu=0}(r)=0$,
and the short-range interaction reduces to the Coulomb interaction,
$w_{ee}^{\sr,\mu=0}(r)=w_{ee}(r)$. Symmetrically, for $\mu\to\infty$,
the short-range interaction vanishes, $w_{ee}^{\sr,\mu\to\infty}(r)=0$,
and the long-range interaction reduces to the Coulomb interaction,
$w_{ee}^{\lr,\mu\to\infty}(r)=w_{ee}(r)$. Physically, $1/\mu$ represents
the distance at which the separation is made.

The Coulombic universal density functional
$F[n]=\min_{\Psi\to n} \bra{\Psi} \hat{T} + \hat{W}_{ee} \ket{\Psi}$~\cite{Levy:79},
where $\hat{T}$ is the kinetic energy operator,
$\hat{W}_{ee} = (1/2) \iint\! d\b{r}_1 d\b{r}_2 w_{ee}(r_{12})
\hat{n}_2(\b{r}_1,\b{r}_2)$ is the Coulomb electron-electron
interaction operator expressed with the pair-density operator $\hat{n}_2(\b{r}_1,\b{r}_2)$,
is then decomposed as
\begin{equation}
 F[n]= F^{\lr,\mu}[n] + E_{\H xc}^{\csr,\mu}[n],
\label{Fdecomp}
\end{equation}
where $F^{\lr,\mu}[n] = \min_{\Psi\to n} \bra{\Psi} \hat{T} +
\hat{W}_{ee}^{\lr,\mu} \ket{\Psi}$ is a long-range universal
density functional associated to the interaction operator
$\hat{W}_{ee}^{\lr,\mu} = (1/2) \iint\! d\b{r}_1 d\b{r}_2
w_{ee}^{\lr,\mu}(r_{12}) \hat{n}_2(\b{r}_1,\b{r}_2)$,
and $E_{\H xc}^{\csr,\mu}[n] = E_{\H}^{\csr,\mu}[n] + E_{xc}^{\csr,\mu}[n]$
is by definition the corresponding complement short-range energy functional,
composed by a trivial short-range Hartree contribution
$E_{\H}^{\csr,\mu}[n] = (1/2) \iint d\b{r}_1 d\b{r}_2
w_{ee}^{\sr,\mu}(r_{12}) n(\b{r}_1) n(\b{r}_2)$
and an unknown short-range exchange-correlation contribution
$E_{xc}^{\csr,\mu}[n]$. At $\mu=0$, the long-range functional reduces to
the usual KS kinetic energy functional, $F^{\lr,\mu=0}[n] =T_s[n]$,
and the short-range functional to the usual Hartree-exchange-correlation functional,
$E_{\H xc}^{\csr,\mu=0}[n] = E_{\H xc}[n]$. In the limit $\mu \to \infty$,
the long-range functional reduces to the Coulombic universal functional,
$F^{\lr,\mu\to\infty}[n] =F[n]$, and the short-range functional vanishes,
$E_{\H xc}^{\csr,\mu\to\infty}[n] = 0$.

The exact ground-state energy of a $N$-electron system in an
external nuclei-electron potential $v_{ne}(\b{r})$,
$E = \min_{n\to N} \Bigl\{ F[n]  + \int\!d\b{r} v_{ne}(\b{r}) n(\b{r})  \Bigl\}$
where the search is over all $N$-representable densities,
can be re-expressed using the long-range/short-range decomposition of $F[n]$
\begin{align}
E &= \min_{n\to N} \Bigl\{ F^{\lr,\mu}[n] + E_{\H xc}^{\csr,\mu}[n]  +
\int\!d\b{r} v_{ne}(\b{r}) n(\b{r})  \Bigl\}
\nonumber\\
  &= \min_{\Psi \to N} \Bigl\{ \bra{\Psi} \hat{T}  +
      \hat{W}_{ee}^{\lr,\mu} \ket{\Psi}
\nonumber \\
  & \qquad \qquad \qquad \!+
      \int\! d\b{r}  v_{ne}(\b{r}) n_{\Psi}(\b{r})  +
      E_{\H xc}^{\csr,\mu}[n_{\Psi}] \Bigl\},
\label{EminPsi}
\end{align}
where the last search is carried out over all $N$-electron normalized
(multi-determinantal) wave functions $\Psi$. In Eq.~(\ref{EminPsi}),
$n_{\Psi}(\b{r})$ is the density coming from the wave function
$\Psi$, i.e. $n_{\Psi}(\b{r})=\bra{\Psi}\hat{n}(\b{r})\ket{\Psi}$ where
$\hat{n}(\b{r})$ is the density operator.

The minimizing wave function $\Psimu$ in Eq.~(\ref{EminPsi})
is given by the corresponding Euler-Lagrange equation
\begin{equation}
  \left( \hat{T} + \hat{W}_{ee}^{\lr,\mu} +
         \hat{V}_{ne} + \hat{V}^{\csr,\mu}_{\H xc}[n_{\Psimu}] \right)
         \ket{\Psimu} =
         \mathcal{E}^{\mu} \ket{\Psimu},
\label{HmuPsi}
\end{equation}
where $\hat{V}_{ne}=\int d\b{r} v_{ne}(\b{r}) \hat{n}(\b{r})$,
$\hat{V}_{\H xc}^{\csr,\mu}[n]\!\!=\!\!\int\!d\b{r}
v^{\csr,\mu}_{\H xc}(\b{r}) \hat{n}(\b{r})$
with the short-range Hartree-exchange-correlation potential
$v_{\H xc}^{\sr,\mu}(\b{r})\! =\! \delta{E}_{\H xc}^{\csr,\mu}[n]/\delta n(\b{r})$,
and $\mathcal{E}^{\mu}$ is the Lagrange multiplier associated
to the constraint of the normalization of the wave function.
Eq.~(\ref{HmuPsi}) defines a long-range interacting effective
Hamiltonian $\hat{H}^{\mu}\!=\!\hat{T}\! +\! \hat{W}_{ee}^{\lr,\mu}\! +\!
\hat{V}_{ne}\! +\!\hat{V}^{\csr,\mu}_{\H xc}[n_{\Psimu}]$
that must be solved iteratively for its multi-determinantal
ground-state wave function $\Psimu$ which gives, in principle,
the exact physical ground-state density
$n(\b{r})=n_{\Psimu}(\b{r})=\bra{\Psimu}\hat{n}(\b{r})\ket{\Psimu}$,
independently of $\mu$. Finally, the exact ground-state energy expression
is thus
\begin{align}
 E = \bra{\Psimu} \hat{T}+\hat{W}_{ee}^{\lr,\mu} +\hat{V}_{ne} \ket{\Psimu} +
 E_{\H xc}^{\csr,\mu}[n_{\Psimu}].
\label{gs_energy}
\end{align}
This exact formalism enables to combine a long-range
wave function calculation with a short-range density functional.
In the special case of $\mu=0$, the KS scheme is recovered,
while the limit $\mu\to\infty$ corresponds to the usual wave function
formulation of the electronic problem.

A short-range LDA~\cite{Toulouse:04b} and other
beyond-LDA~\cite{Toulouse:04a,Toulouse:05a} approximations have
been constructed to successfully describe the functional $E_{xc}^{\csr,\mu}[n]$.
In previous applications of the method, the long-range part of the
calculation has been handled by configuration interaction~\cite{PolSavLeiSto-JCP-02}
or multi-configurational self-consistent field (MCSCF)~\cite{PedJen-JJJ-XX} methods.
We propose in this work to use instead perturbation theory.

\subsection{Range-separated hybrid}
\label{sec:rsh}

At a first level of approximation, we introduce the RSH scheme by
restricting the search in Eq.~(\ref{EminPsi})
to  $N$-electron normalized one-determinant wave functions $\Phi$
\begin{align}
  E^{\mu,\RSH} &= \min_{\Phi \to N} \Bigl\{ \bra{\Phi} \hat{T}  +
      \hat{W}_{ee}^{\lr,\mu} \ket{\Phi}
\nonumber \\
  & \qquad \qquad \qquad \!+
      \int\! d\b{r}  v_{ne}(\b{r}) n_{\Phi}(\b{r})  +
      E_{\H xc}^{\csr,\mu}[n_{\Phi}] \Bigl\}.
\label{EminPhi}
\end{align}

The associated minimizing one-determinant wave function $\Phimu$
satisfies the Euler-Lagrange equation
\begin{equation}
  \left( \hat{T} +
         \hat{V}_{ne} + \hat{V}^{\lr,\mu}_{\H x,\HF}[\Phimu] +
         \hat{V}^{\csr,\mu}_{\H xc}[n_{\Phimu}] \right) \ket{\Phimu} =
         \mathcal{E}^{\mu}_0 \ket{\Phimu},
\label{HmuPhi}
\end{equation}
where $\hat{V}^{\lr,\mu}_{\H x,\HF}[\Phi]$ is a long-range potential operator
appearing due to the restriction to one-determinant wave functions as in
Hartree-Fock (HF) theory, and $\mathcal{E}^{\mu}_0$ is the Lagrange multiplier
associated to the normalization constraint.
As usual, $\hat{V}^{\lr,\mu}_{\H x,\HF}[\Phi]$ is the sum of a Hartree contribution,
$\hat{V}^{\lr,\mu}_{\H,\HF}[\Phi] = \iint d\b{r}_1 d\b{r}_2 w_{ee}^{\lr,\mu}(r_{12})
\bra{\Phi} \hat{n}(\b{r}_1) \ket{\Phi} \hat{n}(\b{r}_2)$,
and a non-local exchange contribution, $\hat{V}^{\lr,\mu}_{x,\HF}[\Phi] =
-(1/2) \iint d\b{r}_1 d\b{r}_2 w_{ee}^{\lr,\mu}(r_{12})
\bra{\Phi} \hat{n}_1(\b{r}_2,\b{r}_1) \ket{\Phi} \hat{n}_1(\b{r}_1,\b{r}_2)$,
where $\hat{n}_1(\b{r}_1,\b{r}_2)$ is the first-order density matrix operator.
Eq.~(\ref{HmuPhi}) defines the RSH non-interacting effective Hamiltonian
$\hat{H}^{\mu}_0 = \hat{T} + \hat{V}_{ne} + \hat{V}^{\lr,\mu}_{\H x,\HF}[\Phimu] +
\hat{V}^{\csr,\mu}_{\H xc}[n_{\Phimu}]$
that must be solved iteratively for its one-determinant ground-state
wave function $\Phimu$. Of course, $\Phimu$ does not give the exact physical density:
$n_{\Phimu}\! \neq\! n$.

The RSH energy expression is finally
\begin{align}
\label{RSH_energy}
  E^{\mu,\RSH} &
   = \bra{\Phimu} \hat{T}+\hat{V}_{ne} \ket{\Phimu} +
  E_{\H x,\HF}^{\lr,\mu}[\Phimu] + E_{\H xc}^{\csr,\mu}[n_{\Phimu}],
\end{align}
where $E_{\H x,\HF}^{\lr,\mu}[\Phi]= \bra{\Phi} \hat{W}_{ee}^{\lr,\mu} \ket{\Phi}$
is the HF-like long-range Hartree-exchange energy.
Eq.~(\ref{RSH_energy}) defines a single-parameter hybrid scheme combining a
long-range HF calculation with a short-range density functional.
The case $\mu=0$ still corresponds to the KS scheme while the method reduces
now to a standard HF calculation in the limit $\mu\to\infty$.


We note that an equivalent to the RSH scheme has been investigated recently by
Pedersen and Jensen~\cite{PedJen-JJJ-XX} as a special case of the combination
of a long-range MCSCF calculation with a short-range density functional.


\subsection{Long-range correlation corrections by perturbation theory}
\label{sec:mp2}

We develop now a long-range perturbation theory, using the RSH determinant
$\Phimu$ as the reference. To do so, we introduce the following energy
expression with a formal coupling constant $\l$
\begin{eqnarray}
  E^{\mu,\l} &=& \min_{\Psi \to N} \Bigl\{ \bra{\Psi} \hat{T}+\hat{V}_{ne}+
  \hat{V}_{\H x,\HF}^{\lr,\mu}[\Phimu] + \l \hat{{\cal W}}^{\lr,\mu} \ket{\Psi}
\nonumber\\
 &&+ E_{\H xc}^{\sr,\mu}[n_\Psi] \Bigl\},
\label{EminPsimul}
\end{eqnarray}
where the search is carried out over all $N$-electron normalized
(multi-determinantal) wave functions $\Psi$ and $\hat{{\cal W}}^{\lr,\mu}$
is the long-range fluctuation potential operator
\begin{equation}
\hat{{\cal W}}^{\lr,\mu} = \hat{W}_{ee}^{\lr,\mu}- \hat{V}_{\H x,\HF}^{\lr,\mu}[\Phimu].
\end{equation}

The minimizing wave function $\Psimul$ in Eq.~(\ref{EminPsimul}) is given
by the Euler-Lagrange equation
\begin{align}
  \left( \hat{T}+\hat{V}_{ne}+  \hat{V}_{\H x,\HF}^{\lr,\mu}[\Phimu] +
  \l \hat{{\cal W}}^{\lr,\mu} + \hat{V}_{\H xc}^{\csr,\mu}[n_{\Psimul}]
  \right) \ket{\Psimul}
\nonumber\\
= \mathcal{E}^{\mu,\l} \ket{\Psimul},
\label{HmulPsi}
\end{align}
where $\mathcal{E}^{\mu,\l}$ is the Lagrange multiplier associated to the
normalization constraint. For $\l=1$, the physical energy is recovered, $E =
E^{\mu,\l=1}$, in principle independently of $\mu$, and Eq.~(\ref{HmulPsi})
reduces to Eq.~(\ref{HmuPsi}): $\Psi^{\mu,\l=1} = \Psimu$,
$\mathcal{E}^{\mu,\l=1}=\mathcal{E}^{\mu}$. For $\l=0$, Eq.~(\ref{HmulPsi})
reduces to the RSH effective Schrödinger equation of Eq.~(\ref{HmuPhi}):
$\Psi^{\mu,\l=0} = \Phimu$, $\mathcal{E}^{\mu,\l=0}=\mathcal{E}^{\mu}_0$.

We expand $E^{\mu,\l}$ in powers of $\l$, $E^{\mu,\l} =
\sum_{k=0}^{\infty} E^{\mu,(k)} \l^k$, and apply the general
results of the non-linear Rayleigh-Schrödinger
perturbation theory~\cite{Angyan:91b,Angyan:93,Gonze:95} outlined
in the Appendix.  It is easy to verify that the sum of zeroth- and first-order
energy contributions gives back the RSH total energy
\begin{equation}\label{eq}
  E^{\mu,(0)}+   E^{\mu,(1)} = E^{\mu,\RSH}.
\end{equation}
The second-order correction can be written as
\begin{eqnarray}\label{Emu2}
E^{\mu,(2)}  &=& -\bra{\Phimu} \hat{{\cal W}}^{\lr,\mu}
   \left( 1+\hat{R}_0^{\mu} \hat{G}_0^{\mu} \right)^{-1}
   \hat{R}_0^{\mu} \hat{{\cal W}}^{\lr,\mu} \ket{\Phimu},
\nonumber\\
\end{eqnarray}
where $\hat{R}_0^{\mu}$ is the reduced resolvent
\begin{eqnarray}
 \hat{R}_0^{\mu} &=& \sum_{I} \frac{ \ket{\Phi^{\mu}_I}
 \bra{\Phi^{\mu}_I}}{{\cal E}_{0,I}^{\mu}-{\cal E}_{0}^{\mu}},
\label{R0mu}
\end{eqnarray}
in terms of the excited eigenfunctions $\Phi^{\mu}_I$ and eigenvalues
$\mathcal{E}_{0,I}^{\mu}$ of the RSH effective Hamiltonian $\hat{H}_{0}^{\mu}$,
and $\hat{G}_0^{\mu}$ is a short-range screening operator
\begin{equation}
  \hat{G}_0^{\sr,\mu} =  2 \iint d\b{r} d\b{r}' \hat{n}(\b{r})\ket{\Phimu}
  f_{\H xc}^{\sr,\mu}[n_{\Phimu}](\b{r},\b{r}') \bra{\Phimu} \hat{n}(\b{r'}),
\end{equation}
with the short-range Hartree-exchange-correlation kernel
$f_{\H xc}^{\sr,\mu}[n](\b{r},\b{r}') =
\delta^2 E_{\H xc}^{\csr,\mu}[n]/\delta n(\b{r}) \delta n(\b{r'})$.

Let insert the spectral resolution of Eq.~(\ref{R0mu}) in Eq.~(\ref{Emu2}).
Since $\hat{\mathcal{W}}^{\lr}$ is a two-electron operator only singly
and doubly excited determinants, $\Phi^{\mu}_{i \to a}$ and $\Phi^{\mu}_{ij \to ab}$
where $i$, $j$ refer to occupied spin-orbitals and $a$, $b$ to virtual spin-orbitals
of $\Phi^{\mu}$, can \textit{a priori} contribute to $E^{\mu,(2)}$.
Actually, singly excited determinants gives vanishing matrix elements,
$ \bra{\Phimu_{i \to a}} \hat{\mathcal{W}}^{\lr} \ket{\Phimu} =0$,
since it can be easily verified that
$\bra{\Phimu_{i \to a}} \hat{W}^{\lr}_{ee} \ket{\Phimu} =
\bra{\Phimu_{i \to a}} \hat{V}_{\H x,\HF}^{\lr,\mu}[\Phimu] \ket{\Phimu}$,
as in standard HF theory. Consequently, the product
$\hat{R}_0^{\mu} \hat{G}_0^{\mu}$ in Eq.~(\ref{Emu2})
involves vanishing matrix elements,
$\bra{\Phimu}\hat{n}(\b{r})\ket{\Phimu_{ij \to ab}}=0$,
i.e. the non-linear terms are zero with the present choice of
the perturbation operator $\hat{\mathcal{W}}^{\lr}$.
The second-order energy correction is thus
\begin{eqnarray}
E^{\mu,(2)} &=& -\bra{\Phimu} \hat{{\cal W}}^{\lr,\mu}
   \hat{R}_0^{\mu} \hat{{\cal W}}^{\lr,\mu} \ket{\Phimu}
\nonumber\\
 &=& \sum_{\substack{i<j\\a<b}}
 \frac{|\bra{\Phimu_{ij \to ab}} \hat{W}_{ee}^{\lr,\mu}
 \ket{\Phimu}|^2 }{ {\cal E}_0^{\mu} - {\cal E}_{0,ij \to ab}^{\mu}}
\nonumber\\
&=& \sum_{\substack{i<j\\a<b}}
  \frac{ \vert \bra{\phi_i^{\mu} \phi_j^{\mu}}
  \hat{w}_{ee}^{\lr,\mu} \ket{\phi_a^{\mu} \phi_b^{\mu}} -
  \bra{\phi_i^{\mu} \phi_j^{\mu}} \hat{w}_{ee}^{\lr,\mu}
  \ket{\phi_b^{\mu} \phi_a^{\mu}} \vert^2}
  {\varepsilon_i^{\mu} + \varepsilon_j^{\mu} -\varepsilon_a^{\mu} - \varepsilon_b^{\mu}},
\nonumber\\
\label{E2}
\end{eqnarray}
where $\phi_k^{\mu}$ is a spin-orbital of $\Phimu$ and
$\varepsilon_k^{\mu}$ is its associated eigenvalue,
$\bra{\phi_i^{\mu} \phi_j^{\mu}} \hat{w}_{ee}^{\lr,\mu} \ket{\phi_a^{\mu} \phi_b^{\mu}}$
are the two-electron integrals associated to the long-range interaction
$w_{ee}^{\lr,\mu}(r_{12})$, and we recall that the indexes $i$, $j$
refer to occupied spin-orbitals and $a$, $b$ to virtual spin-orbitals.
Eq.~(\ref{E2}) is fully analogous to the conventional MP2 energy correction.
The total RSH+MP2 energy is $E^{\mu,\RSH+\MP2} = E^{\mu,\RSH} + E^{\mu,(2)}$.

>From a practical point of view, once the RSH orbitals and one-electron eigenvalues
are available, any standard MP2 implementation can be used, provided that the
long-range electron repulsion integrals corresponding to the RSH orbitals are plugged in.
Due to the long-range nature of these integrals one can take advantage of
efficient modern algorithms, like the local MP2~\cite{Schutz:99},
multipolar integral approximations,
which have particularly favorable convergence properties for long-range
part of the split Coulomb interaction~\cite{Hetzer:00},
or the resolution of identity approach~\cite{Sierka:03}.
It means that in appropriate implementations the extra cost of the MP2 corrections
can be made negligible for large systems with respect to the resolution of the
self-consistent RSH equations, similar to a usual KS calculations with a hybrid functional.
Solid state applications for semi-conductors can also be envisaged on Wannier orbital-based
MP2 implementations~\cite{Pisani:05}.

\section{Results and conclusions}
\label{sec:results}

The above described RSH+MP2 approach has been applied to rare gas dimers, using
a LDA-based short-range exchange-correlation functional with a range-separation parameter of
$\mu\!=\!0.5$. This latter value corresponds to the smallest mean average error of the
atomization energies calculated by  the RSH scheme for the G2-1 set (a subset of 55 molecules
of the G3 set~\cite{Curtiss:97,Curtiss:98}) of small molecules~\cite{Gerber:05}.
This value is in agreement with the intuitive picture predicting that $1/\mu$
should be close to the physical dimensions of a valence electron pair.
The interaction energies were calculated with a
modified version of the MOLPRO package~\cite{molpro-short}. The basis set superposition
error (BSSE) has been removed by the counterpoise method.

The results are presented
as reduced potentials, $U^*(r^*)\!=\! U(r^*\!\cdot\! d_m)/\varepsilon_m$, where
the reduced variables $U^* \!=\! U/\varepsilon_m$ and $r^*\! =\! r/d_m$, are
defined with respect to the equilibrium distance $d_m$ and the well-depth
$\varepsilon_m$ of accurate "experimental" potential curves~\cite{Tang:03}
(cf.~Table~\ref{tab:exp}). The calculated potentials are characterized  by the
hard core radius, $\sigma^*$ defined by $U^*(\sigma^*)\!=\!0$
(experimentally $\sigma^*\!\approx\!0.89$), the reduced well depth, $U_m^*$,
and the  equilibrium distance, $r_m^*$, (experimentally, by construction,
$U_m^* \!=\! -1$ and $r_m^*\!=\!1$). The minimum region is also characterized
by the harmonic vibrational frequencies, $\omega$, related to the
second derivative of the potential at the minimum.

The long-range behaviour of the potential energy curves can be appreciated
from the C$_6$ coefficients.  Experimental  C$_6$ coefficients are
usually obtained from optical data (dipole oscillator strength distributions)
\cite{Kumar:85} and characterize the purely dipolar contribution to the
long-range interaction energy. Since we had no access to such a decomposition
of the interaction energy, we have determined an effective C$_6$ coefficient
by a logarithmic fit of the interaction energies between 30 and 60
Bohrs. This quantity, which includes higher order multipolar effects too,
is presented in the form of a reduced variable,
$\text{C}_6^*\!=\!\text{C}_6/\text{C}_6^{\text{fit}}$. Here
C$_6^\text{fit}$ has been obtained from an analogous fit to the points of
the reference potential reported in Table~\ref{tab:exp}.
For the sake of comparison, the experimental C$_6^\text{exp}$ (purely dipolar)
values are also reported.

\begin{table}
\begin{ruledtabular}
\begin{tabular*}{\hsize}%
{l@{\extracolsep{0ptplus1fil}}c@{\extracolsep{0ptplus1fil}}c@{\extracolsep{0ptplus1fil}}c@{\extracolsep{0ptplus1fil}}c@{\extracolsep{0ptplus1fil}}}
   System    & $d_m$ (a.u.)  &  $\varepsilon_m$ ($\mu$H) &  C$_6^{\text{fit}}$(a.u.) &
   C$_6^{\text{exp}}$(a.u.)  \\
       \colrule
He$_2$       &  5.62    &   ~~34.87    & ~~~1.534         & ~~~1.461   \\
Ne$_2$       &  5.84    &   134.18    & ~~~6.860         & ~~~6.282    \\
Ar$_2$       &  7.10    &   454.50    & ~~73.19~~         & ~~63.75~~    \\
Kr$_2$       &  7.58    &   639.42    & 153.~1~~~         & 129.6~~~    \\
\end{tabular*}
\end{ruledtabular}
\caption{Absolute parameters of the reference potential curves
determined from Ref.~\onlinecite{Tang:03}. The  C$_6^{\text{fit}}$
coefficients were obtained from a logarithmic fit in the same
conditions as explained for the calculated potentials.}
\label{tab:exp}
\end{table}

\begin{figure}
\includegraphics[width=0.97\columnwidth]{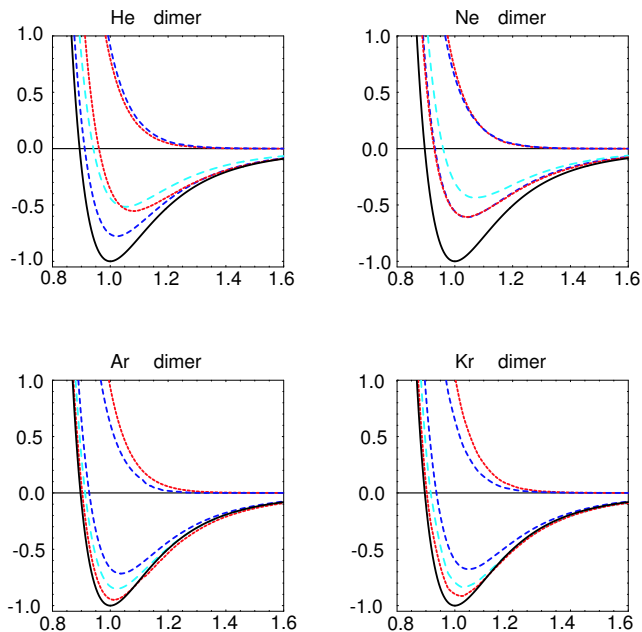} 
\caption{Reduced HF (dotted repulsive), RSH (dashed repulsive),
MP2 (long dashed), CCSD(T) (dashed), RSH+MP2 (dotted)
and Tang-Toennis reference (full) potential curves for
He$_2$, Ne$_2$, Ar$_2$ and Kr$_2$ dimers.}
\label{fig:avtzall}
\end{figure}

The RSH and RSH+MP2 potential curves, as well as the HF, the standard MP2 and the coupled-cluster CCSD(T) ones,
calculated
with the aug-cc-pVTZ basis set are represented for the four dimers
in  Figure~\ref{fig:avtzall}, and compared to the experimental curves.
Note that the reduced representations of
the experimental potentials of different rare gas dimers are practically
indistinguishable. The calculated RSH potentials are always repulsive,
like the HF ones. The RSH+MP2 potentials are slightly too repulsive at short
interatomic distances, as reflected by the values of the hard core radii,
systematically higher than the experiment (around 0.89).
The RSH+MP2 and CCSD(T) curves are almost the same
for Ne$_2$ with a well depth of around $U^*_m\!=\!0.6$, while  the  RSH+MP2 minima
of the Ar$_2$ and  Kr$_2$ systems are even better ($U^*_m\!>\! 0.9$) than the
CCSD(T) ones ($U^*_m\!\approx\! 0.7$).
The position of the minimum is predicted within 1--4\% in the RSH+MP2
approximation. The 6--8\% deviation found for the He$_2$ RSH+MP2
minimum can be explained by an exaggerated repulsion, reflected by the
highest $\sigma^*$ found in this case. In comparison with the usual MP2
potential curves, the RSH+MP2 follow similar trends, being systematically
more stabilizing and closer to the experimental curve.

The main quantitative features of the RSH+MP2 potentials obtained by
the aug-cc-pVTZ and aug-cc-pV5Z basis sets~\cite{Dunning:89,Woon:93,Woon:94,Wilson:99}
are summarized in Table~\ref{tab:avxz_all}
and compared  to the results of  standard MP2 and CCSD(T)
supermolecule calculations with the same basis sets. The
basis set has a non-negligible effect on the calculated
parameters of the potential curves, which converge systematically
towards the experimental values for all the properties. For He$_2$
the double augmented d-aug-cc-pV5Z basis set results are also
included, representing a further improvement of the well depth, but having
practically no effect on the equilibrium distance.

\begin{table*}
\begin{ruledtabular}
\begin{tabular*}{\columnwidth}
{l@{\extracolsep{0ptplus1fil}}c@{\extracolsep{0ptplus1fil}}
c@{\extracolsep{0ptplus1fil}}c@{\extracolsep{0ptplus1fil}}
c@{\extracolsep{0ptplus1fil}}c@{\extracolsep{0ptplus1fil}}
c@{\extracolsep{0ptplus1fil}}
c@{\extracolsep{0ptplus1fil}}c@{\extracolsep{0ptplus1fil}}
c@{\extracolsep{0ptplus1fil}}c@{\extracolsep{0ptplus1fil}}
c@{\extracolsep{0ptplus1fil}}
c@{\extracolsep{0ptplus1fil}}c@{\extracolsep{0ptplus1fil}}
c@{\extracolsep{0ptplus1fil}}c@{\extracolsep{0ptplus1fil}}
c@{\extracolsep{0ptplus1fil}}c@{\extracolsep{0ptplus1fil}}c} Method &
\multicolumn{5}{c}{MP2} & ~~~ &\multicolumn{5}{c}{CCSD(T)} & ~~~
&\multicolumn{5}{c}{RSH+MP2}\\ \cline{2-6}\cline{8-12}\cline{14-18}\
System   & $r_m^*$  & $U_m^*$ & $\omega/\omega_{m}$ & $\sigma^*$ & C$_6^*$ &
~~~      & $r_m^*$  & $U_m^*$ & $\omega/\omega_{m}$ & $\sigma^*$ & C$_6^*$ &
~~~      & $r_m^*$  & $U_m^*$ & $\omega/\omega_{m}$ & $\sigma^*$ & C$_6^*$\\
\colrule
\colrule
         He$_2$  & 1.000   & -1.000   & 1.000   & 0.894 & 1.000 &
              ~~ & 1.000   & -1.000   & 1.000   & 0.894 & 1.000 &
              ~~ & 1.000   & -1.000   & 1.000   & 0.894 & 1.000 \\
         AVTZ    & 1.052   & -0.516   & 1.058   & 0.941 & 0.754  &
              ~~ & 1.023   & -0.777   & 1.047   & 0.912 & 0.966  &
              ~~ & 1.080   & -0.553   & 1.141   & 0.961 & 1.008 \\
         AV5Z    & 1.036   & -0.594   & 1.032   & 0.926 & 0.760  &
              ~~ & 1.007   & -0.896   & 1.019   & 0.896 & 0.980  &
              ~~ & 1.078   & -0.593   & 1.135   & 0.957 & 1.028 \\
          d-AV5Z & 1.032   & -0.629   & 1.032   & 0.923 & 0.763 &
              ~~ & 1.003   & -0.946   & 1.020   & 0.893 & 0.958 &
              ~~ & 1.077   & -0.613   & 1.135   & 0.955 & 1.038 \\
                 &         &          &         &       &       \\
         Ne$_2$  & 1.000   & -1.000   & 1.000   & 0.896 & 1.000 &
              ~~ & 1.000   & -1.000   & 1.000   & 0.896 & 1.000 &
              ~~ & 1.000   & -1.000   & 1.000   & 0.896 & 1.000 \\
         AVTZ    & 1.073   & -0.435   & 1.004   & 0.960 & 0.766 &
              ~~ & 1.041   & -0.609   & 0.983   & 0.931 & 0.914 &
              ~~ & 1.040   & -0.605   & 0.965   & 0.928 & 0.950 \\
         AV5Z    & 1.043   & -0.588   & 0.977   & 0.936 & 0.816 &
              ~~ & 1.009   & -0.877   & 0.950   & 0.904 & 0.990 &
              ~~ & 1.036   & -0.751   & 0.965   & 0.923 & 1.036 \\
                 &         &          &         &       &       \\
         Ar$_2$  & 1.000   & -1.000   & 1.000   & 0.897 & 1.000 &
              ~~ & 1.000   & -1.000   & 1.000   & 0.897 & 1.000 &
              ~~ & 1.000   & -1.000   & 1.000   & 0.897 & 1.000 \\
         AVTZ    & 1.023   & -0.850   & 1.033   & 0.913 & 1.095 &
              ~~ & 1.037   & -0.715   & 1.051   & 0.927 & 0.958 &
              ~~ & 1.012   & -0.948   & 1.013   & 0.903 & 1.154 \\
         AV5Z    & 0.998   & -1.062   & 0.996   & 0.891 & 1.136 &
              ~~ & 1.011   & -0.910   & 1.013   & 0.903 & 1.175  &
              ~~ & 1.007   & -1.040   & 1.003   & 0.896 & 1.215 \\
                 &         &          &         &       &       \\
         Kr$_2$  & 1.000   & -1.000   & 1.000   & 0.896 & 1.000 &
              ~~ & 1.000   & -1.000   & 1.000   & 0.896 & 1.000 &
              ~~ & 1.000   & -1.000   & 1.000   & 0.896 & 1.000 \\
         AVTZ    & 1.029   & -0.840   & 1.049   & 0.918 & 1.124 &
              ~~ & 1.048   & -0.677   & 1.069   & 0.936 & 0.960 &
              ~~ & 1.016   & -0.919   & 1.011   & 0.905 & 1.136 \\
         AV5Z    & 1.002   & -1.080   & 1.021   & 0.893 & 1.153 &
              ~~ & 1.016   & -0.898   & 1.038   & 0.908 & 0.980 &
              ~~ & 1.007   & -1.023   & 1.008   & 0.897 & 1.154 \\
\end{tabular*}
\end{ruledtabular}
\caption{Reduced parameters of the calculated MP2, CCSD(T) and
RSH+MP2 ($\mu\!=$0.5) potential energy curves
obtained by the aug-cc-pVTZ (AVTZ), aug-cc-pV5Z (AV5Z) and d-aug-cc-pV5Z (d-AV5Z) basis sets.
Reduced experimental parameters are listed in the first line for each dimer.
Absolute reference values are given in Table~\ref{tab:exp}.}
\label{tab:avxz_all}
\end{table*}

The basis set superposition error of the equilibrium distances and of the
interaction energies are reported in Table~\ref{tab:bsse}, as the difference
in the parameters of the  BSSE-contaminated and BSSE-free reduced
potential energy curves. The BSSE corrections on the bond lengths and on the
interaction energies are always negative, i.e. the BSSE-contaminated
distances are too short and the energies are too low. In some case, like the
Ne$_2$ dimer with aug-cc-pVTZ basis, the binding energy correction may attain
55 or 67\% of the well depth at the MP2 and CCSD(T) level of approximation.
The corresponding RSH+MP2 BSSE effect is considerably smaller,
but it is still 34\%. The BSSE effect on the bond lengths are much less
spectacular, but still more pronounced in the MP2 and CCSD(T) methods than in
the RSH+MP2 approach. As a general trend we can conclude that the RSH+MP2
has usually less than the half of the MP2 or CCSD(T) basis set superposition
errors. This is a considerable advantage for an efficient and reliable
exploration of potential energy surfaces, especially when the lack of
well-defined subsystems make impossible to perform a counterpoise correction.

\begin{table}
\begin{ruledtabular}
\begin{tabular*}{\hsize}%
{l@{\extracolsep{0ptplus1fil}}c@{\extracolsep{0ptplus1fil}}
c@{\extracolsep{0ptplus1fil}}c@{\extracolsep{0ptplus1fil}}c@{\extracolsep{0ptplus1fil}}
c@{\extracolsep{0ptplus1fil}}c@{\extracolsep{0ptplus1fil}}
c@{\extracolsep{0ptplus1fil}}c@{\extracolsep{0ptplus1fil}}c}
Method & \multicolumn{2}{c}{MP2} & ~ &\multicolumn{2}{c}{CCSD(T)} & ~ &\multicolumn{2}{c}{RSH+MP2}\\
\cline{2-3}\cline{5-6}\cline{8-9}
System & $r_m^*$ & $U_m^*$ & ~ & $r_m^*$ & $U_m^*$ & ~ & $r_m^*$ & $U_m^*$\\
\colrule
He$_2$ &         &         & ~ &         &         & ~ &         &        \\
AVTZ   & -0.007  & -0.125  & ~ & -0.008  & -0.121  & ~ & -0.000  & -0.063 \\
AV5Z   & -0.004  & -0.048  & ~ & -0.002  & -0.037  & ~ & -0.001  & -0.015 \\
d-AV5Z & -0.008  & -0.160  & ~ & -0.007  & -0.113  & ~ & -0.001  & -0.031 \\
\colrule
Ne$_2$ &         &         & ~ &         &         & ~ &         &        \\
AVTZ   & -0.049  & -0.547  & ~ & -0.035  & -0.674  & ~ &  -0.031 & -0.335 \\
AV5Z   & -0.011  & -0.150  & ~ & -0.006  & -0.148  & ~ &  -0.001 & -0.025 \\
\colrule
Ar$_2$ &         &         & ~ &         &         & ~ &         &        \\
AVTZ   & -0.020  & -0.263  & ~ & -0.023  & -0.239  & ~ & -0.007  & -0.101 \\
AV5Z   & -0.005  & -0.138  & ~ & -0.004  & -0.103  & ~ & -0.002  & -0.022 \\
\colrule
Kr$_2$ &         &         & ~ &         &         & ~ &         &        \\
AVTZ   & -0.013 & -0.191   & ~ & -0.017  & -0.174  & ~ & -0.007  & -0.126 \\
AV5Z   & -0.003 & -0.073   & ~ & -0.002  & -0.049  & ~ & -0.002  & -0.039 \\
\end{tabular*}
\end{ruledtabular}
\caption{BSSE correction for the reduced parameters $r_m^*$ and $U_m^*$.}
\label{tab:bsse}
\end{table}

Effective C$_6^*$ coefficients obtained from the RSH+MP2 approach agree
with the experiment within 5\% for He$_2$ and Ne$_2$, and are overestimated by
15--20\% for Ar$_2$ and Kr$_2$. It means that the asymptotic behaviour
of the RSH+MP2 potential curves is reasonable.
We recall that the exact C$_6$ coefficient is given by
the Casimir-Polder relation~\cite{Dobson:02}
\begin{equation}
 \text{C}_6 = \frac{3 \hbar}{\pi} \int_0^\infty d\omega \alpha_1(i\omega)
 \alpha_2(i\omega)
\end{equation}
where $\alpha_1(i\omega)$ and  $\alpha_2(i\omega)$ are the exact
dynamical polarizabilities of the monomers.
It is known that the asymptotic form of the MP2 energy expression corresponds to
an uncoupled HF-type, non-interacting approximation of the monomer
polarizabilities. This means that MP2 calculations do not
reproduce the \textit{exact} C$_6^*$ coefficients: usually they tend to
overestimate them. For instance, in the case of the benzene dimer, this overestimation
in the complete basis limit may reach a factor of 2; for less polarizable
systems the situation is less critical.  An  analogous behaviour is expected for
RSH+MP2. Note however, that in this case one-electron excitations are obtained
from the self-consistent RSH one-electron states, which include, in addition to
the long-range exact exchange,  short-range exchange-correlation effects too.
A more reliable approximation can be developed on the basis of the
adiabatic connection -- fluctuation-dissipation approach~\cite{Kohn:98,Furche:05}
which would ensure, in principle, the exact asymptotic limit of the potential
energy curves. The development of a range-separated version of this method
is under progress.

In conclusion, the RSH+MP2 approach provides an efficient DFT-based
description of weak intermolecular complexes bound by dispersion forces.
Even in its simplest, LDA-based implementation, it represents a
huge improvement over KS calculations, which lead to unreliable
potential curves in the minimum region with a qualitatively wrong asymptotic
behaviour. Range-separated extensions of other density functionals, like the
gradient-corrected PBE functional, are in progress.
By removing systematic errors of currently used approximate
DFT functionals and introducing corrections which grasp the essential physics of
van der Waals interactions, the RSH+MP2 approach extends the applicability of density
functional calculations  to weak intermolecular forces. Further tests
should  decide whether this method is generally  applicable to
the important domains of the  physisorption, or cohesion in
molecular crystals and in layered semi-conductors.

\appendix
\section{Nonlinear Rayleigh-Schrödinger perturbation theory}
\label{sec:nonlinearRSPT}

Let us consider the following general total energy expression,
involving a Hamiltonian $\hat{H}^{(0)}$, a perturbation operator
$\hat{W}$ and a density functional $F[n]$,
\begin{equation}
  E^{\l} = \min_{\Psi \to N} \Bigl\{ \bra{\Psi} \hat{H}^{(0)} +
  \l \hat{W} \ket{\Psi} + F[n_\Psi] \Bigl\},
\label{Elmin}
\end{equation}
where the search is carried out over all $N$-electron normalized
wave functions $\Psi$, $\braket{\Psi}{\Psi} = 1$, and $n_\Psi$ is
the density coming from $\Psi$, $n_\Psi(\b{r}) =\bra{\Psi}\hat{n}(\b{r})\ket{\Psi}$,
where $\hat{n}(\b{r})$ is the density operator. In Eq.~(\ref{Elmin}),
$\l$ is a formal coupling constant; we are ultimately interested
in the case $\l=1$. The minimizing wave function $\Psil$ satisfies
the Euler-Lagrange equation
\begin{equation}
  \left(\hat{H}^{(0)} + \l \hat{W} + \hat{\Omega}^{\l} \right) \ket{\Psil} =
  \mathcal{E}^{\l} \ket{\Psil},
\end{equation}
where the eigenvalue $\mathcal{E}^{\l}$ comes from the normalization
condition and $\hat{\Omega}^{\l}$ is a potential operator coming from
the variation of $F[n]$, non linear in $\l$,
\begin{equation}
 \hat{\Omega}^{\l} = \int d\b{r}
 \frac{\delta F[n^\l]}{\delta n(\b{r})} \hat{n}(\b{r}),
\end{equation}
where $n^\l$ is the density coming from
${\Psil}$, $n_{\l}(\b{r}) =\bra{\Psil}\hat{n}(\b{r})\ket{\Psil}$.

Starting from the reference $\l=0$, we develop a perturbation theory in $\l$.
We introduce the intermediate normalized wave function $\Psitl$
\begin{equation}
\ket{\Psitl} = \frac{\ket{\Psi^{\l}}}{\braket{\Psi^{\l=0}}{\Psi^{\l}}},
\end{equation}
and expand $\Psitl$, $n^{\l}$, $\hat{\Omega}^{\l}$ and
$\mathcal{E}^{\l}$ in powers of $\l$:
$\Psitl = \sum_{k=0}^{\infty} \Psit^{(k)} \l^k$, $n^{\l}=
\sum_{k=0}^{\infty} n^{(k)} \l^k$, $\hat{\Omega}^{\l} =
\sum_{k=0}^{\infty} \hat{\Omega}^{(k)} \l^k$ and $\mathcal{E}^{\l} =
\sum_{k=0}^{\infty} \mathcal{E}^{(k)} \l^k$.
The coefficients $n^{(k)}$ are obtained from the expansion of $\Psitl$ through
\begin{equation}
  n^{\l}(\b{r}) = \frac{ \bra{\Psitl} \hat{n}(\b{r}) \ket{\Psitl} }
  { \braket{\Psitl}{\Psitl}},
\end{equation}
and the coefficients $\hat{\Omega}^{(k)}$ are found from the expansion of $n^{\l}$,
after expanding $\hat{\Omega}^{\l}$ around $n^{(0)}$,
\begin{eqnarray}
 \hat{\Omega}^{\l} &=& \int d\b{r} \frac{\delta F[n^{(0)}]}{\delta n(\b{r})}
 \hat{n}(\b{r})
\nonumber\\
&&+ \iint d\b{r} d\b{r}' \frac{\delta^2 F[n^{(0)}]}{\delta n(\b{r})
\delta n(\b{r}')} \Delta n^{\l}(\b{r}') \hat{n}(\b{r}) + \cdots.
\nonumber\\
\end{eqnarray}
where $\Delta n^{\l}= n^{\l} - n^{(0)}$. The zeroth-order equation is
\begin{equation}
  \left(\hat{H}^{(0)} + \hat{\Omega}^{(0)} \right) \ket{\Psit^{(0)}} =
  \mathcal{E}^{(0)} \ket{\Psit^{(0)}},
\end{equation}
and of course $\Psit^{(0)} = \Psi^{\l=0}$. For the general order $k \geq 1$,
\begin{align}
\left( \hat{H}^{(0)} + \hat{\Omega}^{(0)}-\mathcal{E}^{(0)} \right) \ket{\Psit^{(k)}}
+ \hat{W}\ket{\Psit^{(k-1)}}
\nonumber\\
+ \sum_{i=1}^{k}
  \hat{\Omega}^{(i)}\ket{\Psit^{(k-i)}} =
  \sum_{i=0}^{k}\mathcal{E}^{(i)}
  \ket{\Psit^{(k-i)}}.
\end{align}
The corresponding eigenvalue correction of order $k$ is
\begin{equation}
 \mathcal{E}^{(k)} = \bra{\Psit^{(0)}}\hat{W}\ket{\Psit^{(k-1)}}
 + \sum_{i=1}^k  \bra{\Psit^{(0)}}\hat{\Omega}^{(i)}\ket{\Psit^{(k-i)}},
\end{equation}
containing, besides the usual first term, a "non-linearity" term as well.
Introducing the reduced resolvent, $\hat{R}_0$,
\begin{eqnarray}
\hat{R}_0 &=& \sum_{I} \frac{ \ket{\Psit^{(0)}_I}
\bra{\Psit^{(0)}_I} }{\mathcal{E}_I^{(0)}-\mathcal{E}^{(0)}},
\label{R0}
\end{eqnarray}
where $\Psit^{(0)}_I$ and $\mathcal{E}^{(0)}_I$ are the excited eigenfunctions
and eigenvalues of $\hat{H}^{(0)}$, the wave function correction of order $k$ writes
\begin{align}
\ket{\Psit^{(k)}}  =
  - \hat{R}_0\hat{W}\ket{\Psit^{(k-1)}} -
   \hat{R}_0\hat{\Omega}^{(k)}\ket{\Psit^{(0)}}
\nonumber\\
  - \hat{R}_0\sum_{i=1}^{k-1}
  \left(\hat{\Omega}^{(i)}-\mathcal{E}^{(i)}\right)\ket{\Psit^{(k-i)}}.
\end{align}

The total energy can be re-expressed in terms of the eigenvalue
$\mathcal{E}^{\l}$ and the "double counting correction" $D^{\l}$
\begin{equation}
  E^{\l} = \mathcal{E}^{\l} + D^{\l},
\end{equation}
where
\begin{equation}
  D^{\l} =  F[n^{\l}]
 - \int  d\b{r} \frac{\delta F[n^\l]}{\delta n(\b{r})} n^{\l}(\b{r}).
\end{equation}
We expand $E^{\l}$ and $D^{\l}$ in powers of $\l$: $E^{\l} =
\sum_{k=0}^{\infty} E^{(k)} \l^k$ and $D^{\l} = \sum_{k=0}^{\infty} D^{(k)} \l^k$.
The coefficients $D^{(k)}$ are found from the expansion of $n^{\l}$,
after expanding $D^{\l}$ around $n^{(0)}$,
\begin{eqnarray}
  D^{\l} &=&  F[n^{(0)}] + \int  d\b{r}
  \frac{\delta F[n^{(0)}]}{\delta n(\b{r})}  \Delta n^{\l}(\b{r})
\nonumber\\
&& + \frac{1}{2}\iint d\b{r} d\b{r}'
\frac{\delta^2 F[n^{(0)}]}{\delta n(\b{r}) \delta n(\b{r}')} \Delta n^{\l}(\b{r}')
\Delta n^{\l}(\b{r}) + \cdots
\nonumber\\
 && - \int  d\b{r} \frac{\delta F[n^{(0)}]}{\delta n(\b{r})} n^{\l}(\b{r})
\nonumber\\
 &&- \iint  d\b{r} d\b{r}'  \frac{\delta^2 F[n^{(0)}]}{\delta n(\b{r})
 \delta n(\b{r}')} \Delta n^{\l}(\b{r}') n^{\l}(\b{r}) - \cdots.
\nonumber\\
\end{eqnarray}
The zeroth-order total energy is simply
\begin{equation}\label{eq}
     E^{(0)}  = \mathcal{E}^{(0)} + F[n^{(0)}] - \int d\b{r}
     \frac{\delta F[n^{(0)}]}{\delta n(\b{r})} n^{(0)}(\b{r}),
\end{equation}
The general correction of order $k\geq 1$ writes
\begin{equation}
  E^{(k)} = \bra{\Psit^{(0)}}\hat{W}\ket{\Psit^{(k-1)}} + \Delta^{(k)}
\end{equation}
where $\Delta^{(k)}$ is
\begin{equation}
 \Delta^{(k)} = \sum_{i=1}^k  \bra{\Psit^{(0)}}
 \hat{\Omega}^{(i)}\ket{\Psit^{(k-i)}} + D^{(k)} .
\end{equation}
At first order, it can be verified that the nonlinearity term of the
eigenvalue and the double counting correction cancel each other,
i.e. $\Delta^{(1)}=0$, and we obtain the conventional first-order energy correction
\begin{align}
E^{(1)}  = \bra{\Psit^{(0)}}\hat{W}\ket{\Psit^{(0)}}.
\end{align}
At second order, the situation is analogous, i.e. $\Delta^{(2)}=0$,
and again the conventional form of the energy correction is retrieved
\begin{align}
   E^{(2)}  = \bra{\Psit^{(0)}}\hat{W}\ket{\Psit^{(1)}}.
\end{align}
The nonlinearity effects are "hidden" in the first-order
wave function correction, which can be obtained from the self-consistent
equation:
\begin{align}
  \ket{\Psit^{(1)}}  =
  - \hat{R}_0\hat{W}\ket{\Psit^{(0)}} -
   \hat{R}_0\hat{\Omega}^{(1)}\ket{\Psit^{(0)}}
\label{Psit1}
\end{align}
Since the first-order potential operator is, for real wave functions,
\begin{equation}\label{eq}
\hat{\Omega}^{(1)} =  2 \iint d\b{r} d\b{r}' \frac{\delta^2
F[n^{(0)}]}{\delta n(\b{r}) \delta n(\b{r}')}
\bra{\Psit^{(0)}}\hat{n}(\b{r}')\ket{\Psit^{(1)}} \hat{n}(\b{r}),
\end{equation}
Eq.~(\ref{Psit1}) can be re-expressed as
\begin{align}\label{eq}
  \ket{\Psit^{(1)}}  =
  - \hat{R}_0\hat{W}\ket{\Psit^{(0)}} -
   \hat{R}_0 \hat{G}_0 \ket{\Psit^{(1)}},
\end{align}
where
\begin{equation}\label{eq}
  \hat{G}_0 =  2 \iint d\b{r} d\b{r}' \hat{n}(\b{r})\ket{\Psit^{(0)}}
  \frac{\delta^2
F[n^{(0)}]}{\delta n(\b{r}) \delta n(\b{r}')} \bra{\Psit^{(0)}} \hat{n}(\b{r'}).
\end{equation}
The final expression of the second-order energy correction can be written as
the series
\begin{eqnarray}\label{eq}
   E^{(2)}  &=& -\bra{\Psit^{(0)}}\hat{W}
   \bigl(1+\hat{R}_0 \hat{G}_0\bigr)^{-1}\hat{R}_0\hat{W}\ket{\Psit^{(0)}}
\nonumber\\
   &=& -\bra{\Psit^{(0)}}\hat{W}\hat{R}_0\hat{W}\ket{\Psit^{(0)}}
\nonumber\\
&& +
  \bra{\Psit^{(0)}}\hat{W}\hat{R}_0 \hat{G}_0\hat{R}_0\hat{W}\ket{\Psit^{(0)}}
\nonumber\\
  &&- \cdots.
\end{eqnarray}
Further details and higher-order expressions will be given in a
forthcoming publication.


\end{document}